\documentclass[aps,prl,amsmath,amssymb,twocolumn,superscriptaddress]{revtex4-2}
\usepackage[english]{babel}
\usepackage[utf8]{inputenc}
\usepackage{amsfonts}
\usepackage[T1]{fontenc}
\usepackage[pdftex]{graphicx}
\usepackage{fancyhdr}
\usepackage{hyperref}
\usepackage{url}
\usepackage[titletoc,title]{appendix}
\usepackage{epstopdf}
\usepackage{amsmath}
\usepackage{amssymb}
\usepackage{dsfont}
\usepackage{mathtools}
\usepackage{bbold}
\usepackage{xcolor}
\usepackage{tikz}
\usepackage{cancel}
\usepackage{soul}
\usetikzlibrary{arrows.meta}

\hypersetup{
    bookmarks=false,         
    unicode=false,          
    pdftoolbar=false,        
    pdfmenubar=true,        
    pdffitwindow=false,     
    pdfstartview={FitH},    
    pdftitle={Probing the many-body localized spin-glass phase through quench dynamics},    
    pdfauthor={P. Brighi et al.},     
    pdfsubject={},   
    pdfcreator={},   
    pdfproducer={}, 
    pdfkeywords={},
    pdfnewwindow=true,      
    colorlinks=true,       
    linkcolor=black,          
    citecolor=blue,        
    filecolor=magenta,      
    urlcolor=blue           
}

\graphicspath{{./Figures/}}

\newcommand{\tr}{{\rm tr}}

\newcommand{\ket}[1]{|#1\rangle}
\newcommand{\bra}[1]{\langle#1|}

\newcommand{\sz}[1]{S^z_{#1}}
\newcommand{\sx}[1]{S^x_{#1}}

\newcommand{\mytitle}{Probing the many-body localized spin-glass phase through quench dynamics}

\begin{document}

\title{\mytitle}

\author{Pietro Brighi}
\affiliation{Faculty of Physics, University of Vienna, Boltzmanngasse 5, 1090 Vienna, Austria}
\author{Marko Ljubotina}
\affiliation{Institute of Science and Technology Austria, am Campus 1, 3400 Klosterneuburg, Austria}
\affiliation{Technical University of Munich, TUM School of Natural Sciences, Physics Department, 85748 Garching, Germany}
\affiliation{Munich Center for Quantum Science and Technology (MCQST), Schellingstr. 4, München 80799, Germany}
\author{Maksym Serbyn}
\affiliation{Institute of Science and Technology Austria, am Campus 1, 3400 Klosterneuburg, Austria}
\date{\today}

\begin{abstract}
Eigenstates of quantum many-body systems are often used to define phases of matter in and out of equilibrium; however, experimentally accessing highly excited eigenstates is a challenging task, calling for alternative strategies to dynamically probe nonequilibrium phases.
In this work, we characterize the dynamical properties of a disordered spin chain, focusing on the spin-glass regime.
Using tensor-network simulations, we observe oscillatory behavior of local expectation values and bipartite entanglement entropy.
We explain these oscillations deep in the many-body localized spin glass regime via a simple theoretical model.
From perturbation theory, we predict the timescales up to which our analytical description is valid and confirm it with numerical simulations.
Finally, we study the correlation length dynamics, which, after a long-time plateau, resumes growing in line with renormalization group (RG) expectations.
Our work suggests that RG predictions can be quantitatively tested against numerical simulations and experiments, potentially enabling microscopic descriptions of dynamical phases in large systems.
\end{abstract}

\maketitle

\noindent\textit{Introduction.---}The fate of interacting quantum systems evolving under unitary dynamics is an outstanding question in modern quantum many-body physics that has received considerable attention in recent years, both theoretically~\cite{Rigol2007,Rigol2008,Rigol2009,Calabrese2014,Fagotti2016, D'alessio2016} and experimentally~\cite{Weiss2006,Bloch2008,Bloch2012,Schmiedmayer2012,Rosch2012,Nagerl2013,Lukin2017}. 
Generic systems are believed to follow the eigenstate thermalization hypothesis (ETH)~\cite{Deutsch1991,Srednicki1994}, resulting in thermal expectation values for local observables at late times.
On the other hand, integrable~\cite{Sutherland2004} and disordered models show a breakdown of ergodicity and absence of thermalization.
While integrability is extremely fragile to perturbations, many-body localized systems~\cite{Basko2006,Gornyi2005,Huse2015,Serbyn2019,Sierant2024} provide a robust way of escaping thermalization on experimentally relevant sizes and timescales~\cite{Schreiber2015,Choi2016,Bloch2017,Greiner2019,Wang2021}.

More recently, increasing attention has been devoted to the interplay of disorder and topological order~\cite{Pollmann2014,Huse2020,Yao2021,Laflorencie2022,Beri2022,Orito2022,Pal2023,Laflorencie2023,Beri2024,Laflorencie2024}, as shown e.g.~in the interacting Majorana chain.
In these systems, the disorder of the local potential and of the bond coupling compete, and two distinct MBL phases emerge~\cite{Huse2013,Pollmann2014,Huse2020,Yao2021,Laflorencie2022}.
When the on-site disorder dominates, the system is in a paramagnetic MBL phase~\cite{Laflorencie2022,Huse2020,Yao2021}, where the original topological order is completely destroyed and the system behaves according to standard MBL: area-law entanglement of eigenstates~\cite{Serbyn2013,Bauer2013}, Poissonian level statistics~\cite{Oganesyan2007}, and slow entanglement growth~\cite{Bardarson2012}.
However, in the opposite scenario of dominating bond disorder the system enters a novel MBL phase, characterized by long-range correlations, degeneracies in the spectrum and a cat-like structure of the eigenstates~\cite{Pollmann2014,Laflorencie2022,Huse2020,Yao2021} leading to higher entanglement as compared to the paramagnet MBL.
In spite of the growing interest, the dynamical characterization of this MBL spin-glass (MBL-SG) phase is still limited to small systems~\cite{Orito2022}.

In this work, we study the dynamics of random product states in the extended Ising model introduced in Refs.~\cite{Huse2020,Yao2021,Laflorencie2022}.
In particular, we focus on the dynamics deep in the MBL-SG phase.
We show that this regime is characterized by oscillatory dynamics both in local observables and entanglement entropy.
Through an approximate description of the time evolution, we analytically provide accurate predictions for the behavior of magnetization and entanglement.
We show that the local magnetization oscillates with frequencies depending on the bond disorder and on the interaction strength.
We further explain the bounds on entanglement growth and its oscillations through the matrix-product-operator (MPO) representation of the approximate time propagator~\cite{DeNicola2021}.
Finally, we extract the correlation length from the exponential decay of the correlation function.
Our study shows that the correlation length initially saturates to a finite value, but resumes growing on a timescale scaling exponentially with the disorder strength.
The unbounded growth of the correlation length is in line with the long-range order detected in highly excited eigenstates within the MBL-SG phase.

\noindent\textit{Model.---}We study an extended Ising model with random couplings $J_i\in[0,W_J]$ and random transverse fields $h_i\in[0,W_h]$
\begin{equation}
\label{Eq:H}
H \!=\! \sum_i\! J_i\sx{i}\sx{i+1} \!+\! h_i\sz{i} \!+\! \lambda(W_J\sx{i}\sx{i+2} \!+\! W_h\sz{i}\sz{i+1}),
\end{equation}
where  spin-1/2 operators are expressed via Pauli matrices, $S^\alpha_j = \sigma^\alpha_j/2$. In this model the interaction strength $\lambda$ sets the scale for the next-nearest-neighbor Ising interaction and for the coupling along the $z$ direction. 
Throughout this paper, we fix the interaction strength to $\lambda = 0.5$, and we define a disorder parameter $\delta = \overline{\log{J_i}} - \overline{\log{h_i}}\approx\log{({W_J}/{W_h})}$, where the bar indicates average over disorder realizations and lattice sites.
To avoid introducing large energy scales, we fix $\max(W_J,W_h) = 1$.

The model has $\mathbb{Z}_2$ symmetry, given by $M=\prod_i \sigma^z_i$.
Additionally, the Hamiltonian~(\ref{Eq:H}) is self-dual under the transformation $\tau^z_i = \sigma^x_i\sigma^x_{i+1}$, $\tau^x_i = \prod_{j\leq i} \sigma^z_j$ and $\delta\to-\delta$~\cite{Orito2022}.
The Hamiltonian~(\ref{Eq:H}) can further be mapped to an interacting Majorana chain through a Jordan-Wigner transformation to spinless fermions and an additional introduction of Majorana fermions~\cite{Laflorencie2022}.

Previous exact diagonalization (ED) studies of model~(\ref{Eq:H}) have shown the emergence of three distinct phases~\cite{Huse2013,Pollmann2014,Laflorencie2022,Huse2020,Yao2021}.
At large negative $\delta$, where disorder in the local magnetic field is dominant, there exists a localized paramagnetic phase presenting the characteristic features of standard MBL phases.
At intermediate $\delta$ around $0$, the model enters an ergodic phase, characterized by volume-law entanglement entropy of the eigenstates~\cite{Huse2020,Yao2021,Laflorencie2022}.
Finally, at large positive $\delta$ there is a transition to the MBL spin-glass phase.
This latter phase has interesting remnant topological features, such as spectral pairing and long-range order in the magnetization~\cite{Huse2013,Pollmann2014,Huse2020,Yao2021,Laflorencie2022}.

In the remainder of this work we will investigate the dynamics deep in the spin-glass phase, where we can provide an analytical theory explaining our numerical observations. 
We will focus on the dynamics from random product states in the computational basis that can be easily prepared and probed experimentally in a variety of quantum simulation platforms~\cite{Lukin2017,Bloch2017Rev}.
Unless otherwise specified, the data reported are obtained averaging over $50$ random initial states, each corresponding to a different disorder realization.

\begin{figure}[b]
    \centering
    \includegraphics[width=0.99\linewidth]{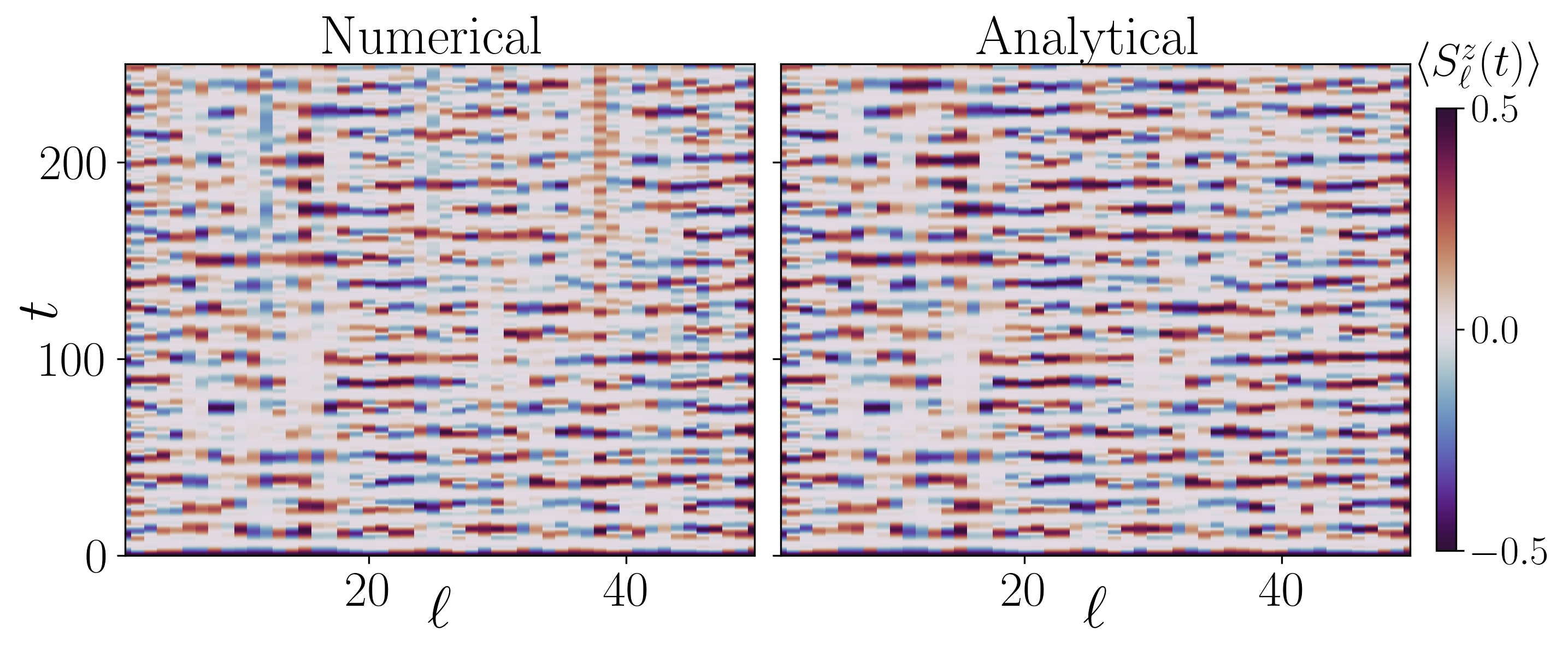}
    \caption{\label{Fig:Sz}
    Local magnetization dynamics for a fixed disorder realization at $\delta = 4$. 
    Comparison of analytical results obtained from Eq.~(\ref{Eq:szt}) with numerical simulations with bond dimension $\chi=128$ shows quantitative agreement up to times $\sim 100$ and captures the main qualitative features of dynamics even at longer times. 
    }
\end{figure}

\noindent\textit{Magnetization dynamics.---}First, we analyze the local magnetization dynamics $\langle\sz{\ell}(t)\rangle = \bra{\psi_0}e^{\imath Ht}\sz{\ell}e^{-\imath Ht}\ket{\psi_0}$.
While in the MBL paramagnetic phase the strong local fields favor freezing of the local magnetization close to its initial value, at strong positive $\delta$ bond disorder is dominant and magnetization is expected to vanish~\cite{Altman2013,Zakrzewski2024}.
However, we find that the decay of magnetization is not entirely featureless and carries information about the local disorder in the long-lived oscillations that characterize its dynamics.

\begin{figure}[tb]
    \centering
    \includegraphics[width=0.99\linewidth]{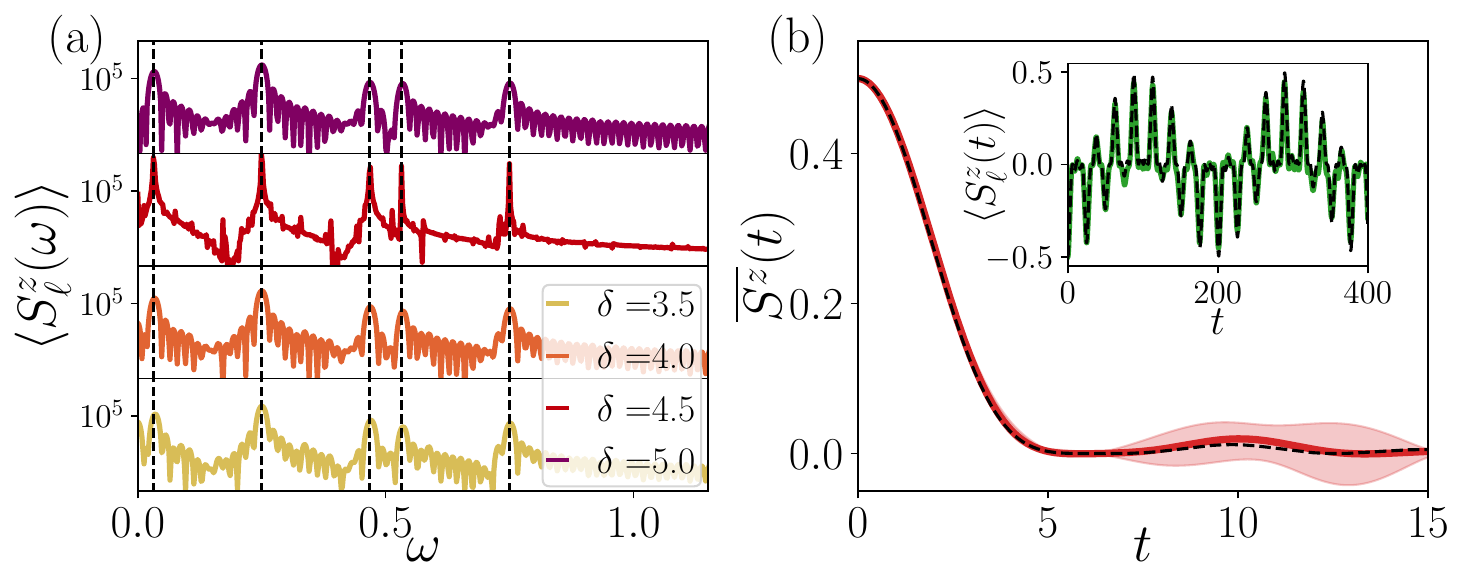}
    \caption{\label{Fig:Sz Fourier + global}
    (a):~The Fourier transform of the local magnetization for a single disorder realization $\langle\sz{\ell}(\omega)\rangle$ shows exactly six peaks whose frequency is set by the local random couplings and by the interaction as predicted by the expressions $\omega_{j}$ (black dashed lines).
    The peaks do not move as the value of $\delta$ is changed, in agreement with our convention of fixing $W_J=1$ when $\delta>0$.
    (b):~The average global magnetization $\overline{\sz{}}(t)$ at $\delta=4$ shows a quadratic decay in agreement with the predictions of Eq.~(\ref{Eq:sz average}) (dashed black line).
    At later times, the oscillations around $0$ are still compatible with the analytical prediction within the uncertainty due to finite disorder average (shaded area).
    In the inset, we further show the good agreement over long times of the single local magnetization dynamics with the analytical expression Eq.~(\ref{Eq:szt}) (black dashed line).
    Data were obtained through TEBD with bond dimension $\chi = 128$.
    }
\end{figure}

\begin{figure*}
    \centering
    \includegraphics[width=0.99\textwidth]{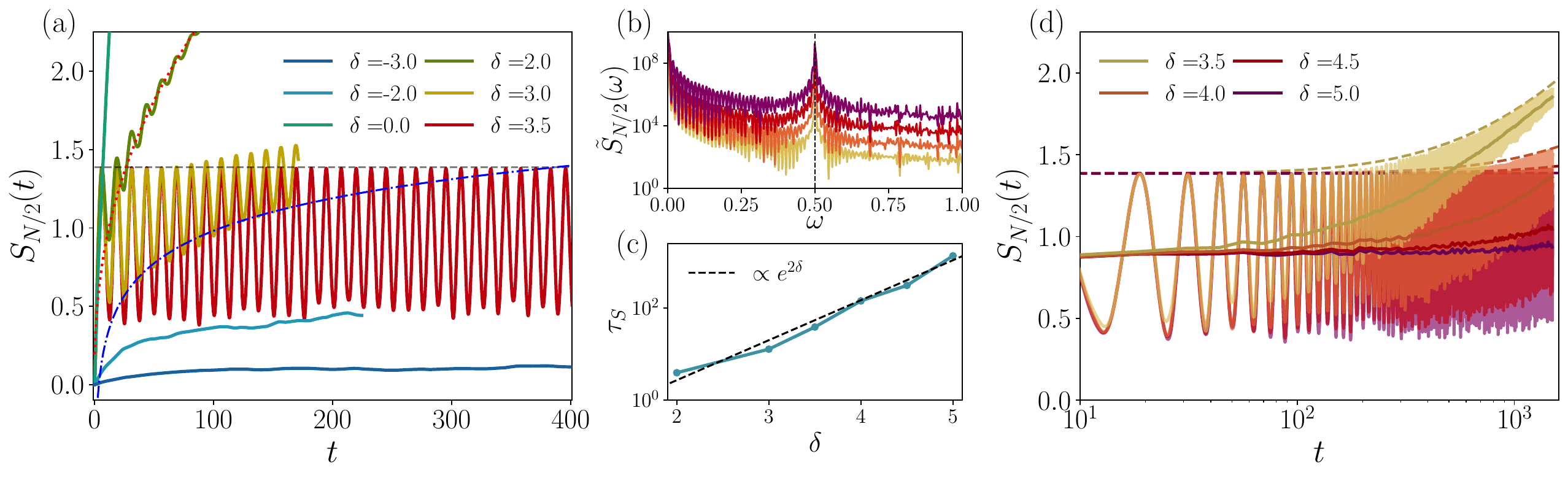}
    \caption{\label{Fig:S}
    (a): The half-chain entanglement entropy $S_{N/2}(t)$ exhibits slow logarithmic growth in the MBL paramagnet phase ($\delta = -3,-2$), as opposed to the fast growth shown in the ergodic pase $\delta = 0$.
    In the spin-glass regime ($\delta > 0$), instead, $S(t)$ features distinctive oscillations, with amplitude bounded from above by $\log(4)$ (gray dashed line) up to a timescale $\tau_S(\delta)$.
    When these oscillations fade out, $S(t)$ grows as power-law (dashed red line $\sim t^{0.42}$) in the ergodic case $\delta = 2$ and as logarithm for $\delta = 3$ (dashed blue line $\sim 0.3\log(t)$).
    (b): The Fourier transform of the entanglement entropy $\tilde{S}_{N/2}(\omega)$ highlights a single peak at frequency $\omega_0 = \lambda W_J = 1/2$, as analytically predicted.
    (c): We numerically obtain the timescale $\tau_S(\delta)$ when the approximation for $U_t$ breaks down from the fit to the square logarithmic growth predicted by renormalization group.
    Its scaling is consistent with our theoretical estimate $e^{2\delta}$. 
    (d): At long times, entanglement eventually starts to grow.
    We show both the instantaneous entanglement entropy with its oscillations (shaded curves) and its average over each oscillation period $[nT,(n+1)T]$, with $T=2\pi/\omega_0=4\pi$(solid curves).
    The upper envelope of the entanglement entropy is compatible with a $\log^2(t)$ growth (dashed lines), as predicted by RG calculations in Ref.~\cite{Altman2013}.
    }
\end{figure*}

Deep in the spin-glass phase, $\delta\gg1$, one can neglect the terms containing the $S^z$ operators in the Hamiltonian, resulting in the approximate time propagator
\begin{equation}
    \label{Eq:Ut}
    U_t = e^{-\imath Ht} \approx e^{-\imath\left( \sum_i J_i\sx{i}\sx{i+1} + \lambda W_J\sx{i}\sx{i+2} \right)t}.
\end{equation}
Using this approximation, we analytically obtain $\langle \sz{\ell}(t)\rangle$ for each individual disorder realization~\cite{Supplementary} as
\begin{equation}
    \label{Eq:szt}
    \langle \sz{\ell}(t)\rangle = \langle\sz{\ell}\rangle_0 \cos \frac{J_\ell t}{2}\cos\frac{J_{\ell-1}t}{2} \cos^2\frac{W_J\lambda t}{2} ,
\end{equation}
where $\langle \sz{\ell}\rangle_0 = \pm 1/2$ is the initial value of the magnetization on site $\ell$.
Knowing all disordered couplings we obtain the dynamics of the local magnetization in the system and compare it with our numerical results obtained via time-evolving-block-decimation (TEBD)~\cite{Vidal2003} with bond dimension $\chi = 128$ \cite{Supplementary}.
As we show in Fig.~\ref{Fig:Sz} and in the inset of Fig.~\ref{Fig:Sz Fourier + global}(b), our analytical prediction captures the quantitative behavior up to times $t\sim 100$, and continues to describe the dynamics qualitatively at even longer timescales. 

This theoretical result provides a reliable scheme to understand dynamical behaviors in the MBL-SG phase. In particular, Eq.~(\ref{Eq:szt}) predicts the emergence of oscillations in the magnetization dynamics, with $6$ resonant frequencies $\omega_{j}$.
The frequencies $\omega_j$ can be obtained from the Fourier transform of the expression in Eq.~(\ref{Eq:szt}) $\langle\sz{\ell}(\omega)\rangle = \int dt e^{\imath\omega t} \langle\sz{\ell}(t)\rangle$, yielding $\omega_{1,\ldots,4} = \frac{1}{2}|J_\ell \pm J_{\ell-1} \pm 2\lambda|$, 
$\omega_{5,6} = \frac{1}{2}|J_\ell \pm J_{\ell-1} |$.
In Figure~\ref{Fig:Sz Fourier + global}(a) we report the Fourier transform of the local magnetization at a given site for a single disorder realization and for different values of $\delta$ within the spin-glass phase.
The Fourier signal shows six evident peaks at the frequencies predicted by our analytical results (black dashed lines), with no dependence on the disorder strength $\delta$.

The analytical expression for $\langle \sz{\ell}(t)\rangle$ obtained above further allows to calculate exact disorder averages.
Indeed, each disorder instance $(n)$ produces an independent set of random frequencies determined by $J^{(n)}_{\ell}$ and $J^{(n)}_{\ell-1}$, and the disorder average can be obtained by integrating over the distribution function $f(J) = \theta(1-J)\theta(J)$, where $\theta(x)$ is Heaviside's theta function.
The average local magnetization then reads
\begin{equation}
    \label{Eq:sz average}
    \overline{\langle\sz{\ell}(t)\rangle} = 4\langle\sz{\ell}\rangle_0\frac{\cos^2\left( \frac{\lambda W_J}{2}t \right)\sin^2\left(\frac{W_J}{2} t\right)}{t^2}.
\end{equation}
Defining the average magnetization for the $N_+$ sites where $\langle\sz{\ell}\rangle_0 = 1/2$ as $S^z_+ $ and analogously $S^z_-$, we obtain the global average magnetization as $\overline{\sz{}}(t) = (S^z_+ - S^z_-)/N = (4/t^2)\cos^2\left( \frac{\lambda W_J}{2}t \right)\sin^2\left(\frac{W_J}{2} t\right)$, analogous to the imbalance introduced in N\'eel initial states~\cite{Schreiber2015,Bloch2017}.
The exact disorder average shows a fast power law decay of the global magnetization, as expected due to the dominant bond disorder in the MBL spin-glass regime. 
This can be contrasted to the RG study of the random coupling spin-1/2 XXZ chain~\cite{Altman2013}, where the Ne\'el order parameter decays parametrically slower. 

In Figure~\ref{Fig:Sz Fourier + global}(b) we compare the prediction of Eq.~(\ref{Eq:sz average}) with numerical simulations.
The short time decay of $\overline{\sz{}}$ agrees well with our analytic expression, confirming the fast decay $\sim1/t^2$.
At longer times, numerical results still show small oscillations, but within uncertainty their value agrees with the exact disorder average decaying monotonically. 
In summary our theoretical description accurately captures both the single realization oscillations and the average decay of the local magnetization, thus indicating measurable dynamical features of the MBL-SG phase.

\noindent\textit{Entanglement dynamics.---}Next, we study the behavior of the half-chain entanglement entropy, defined as the von Neumann entropy of the reduced density matrix $\rho_A(t) = \tr_B \rho(t)$, where $A$ and $B$ are the two halves of the chain, $S_{N/2}(t) = -\tr \rho_A(t)\log\rho_A(t)$.
Entanglement entropy is commonly used to dynamically distinguish ergodic phases of matter, where it grows algebraically in time~\cite{Calabrese2005,Huse2013a}, from many-body localized ones where this growth is only logarithmic~\cite{Znidaric2008,Bardarson2012}.
Here, we find that the MBL spin-glass phase presents a different behavior, which markedly distinguishes it from the standard MBL paramagnetic phase at negative $\delta$.

The entanglement dynamics, presented in Figure~\ref{Fig:S}(a), show three clearly distinct behaviors in the different phases of the model.
In the MBL paramagnetic phase ($\delta \lesssim -2$) the entanglement entropy grows logarithmically in time, as expected for strongly disordered interacting systems~\cite{Znidaric2008,Bardarson2012}.
As $\delta$ approaches zero, the system enters the ergodic regime, where entanglement grows as a power law, limiting our numerical simulations.
Increasing $\delta$ further, the MBL spin-glass regime sets in, where entanglement shows clear oscillatory behavior.
These oscillations eventually flatten out, as one can observe for $\delta=2$, and make way to a more standard entanglement growth.

Similarly to the local magnetization case, we use the approximate time-evolution operator, Eq.~(\ref{Eq:Ut}), to explain the main features observed in this regime.
Following Ref.~\cite{DeNicola2021}, we write $U_t$ as a bond dimension $\chi=4$ matrix-product-operator (MPO), whose entries oscillate at a frequency $\omega_0 + \delta\omega$~\cite{Supplementary}.
While $\delta\omega$ depends on the disorder realization and averages out, the main frequency $\omega_0 = \lambda W_J$ is common to all realizations and governs the oscillations observed in Fig.~\ref{Fig:S}(a).
We further confirm this in Figure~\ref{Fig:S}(b) by comparing the Fourier transform of the entanglement entropy with the analytical expression $\omega_0$.
The numerical results clearly show a well defined peak at the predicted value, marked by a black dashed line.
As in the MBL spin-glass phase we fix $W_J=1$, the position of the peak does not depend on $\delta$.

The small bond dimension of the MPO representation of the time-propagator limits the growth of entanglement entropy to $S(t)\leq \log(\chi) = \log(4)$.
This is clearly visible in the numerical results shown in Figure~\ref{Fig:S}(a), where the curves at $\delta>0$ remain below this value (dashed gray line) up to a time $\tau$ increasing with~$\delta$.
The growth of $S(t)$ above $\log(4)$ is due to an increase in the bond dimension of the MPO representation of $U_t$, and therefore to the breakdown of the approximation neglecting the perturbative part of the Hamiltonian $H_{W_h} = h_i\sz{i} + \lambda W_h \sum_i \sz{i}\sz{i+1}$.
Inspired by Fermi's golden rule, we use the square of the perturbation strength to estimate the timescale at which its effects become relevant, $\tau_S\approx\langle H_{W_h}\rangle^{-2}\approx 1/W_h^2 = e^{2\delta}$.
Figure~\ref{Fig:S}(c) compares this result to $\tau_S(\delta)$ obtained numerically, revealing an accurate agreement.

Beyond this timescale, entanglement starts increasing. 
While for low values of $\delta$ this growth corresponds to a power law in time (e.g.~the red dotted curve for $\delta=2$), at higher disorders the behavior is in line with predictions for bond disordered MBL.
In particular, we notice that the lower envelope of the oscillation grows logarithmically in time (blue dotted curve), while the upper one is well captured by $\log^2(t)$, as shown by the dashed lines in panel (d).
This suggests that the time-averaged entanglement entropy [solid lines in panel (d)] grows as $\log^\alpha(t)$.
Although the timescales we are able to reach are too short to determine the power $\alpha$ accurately and we cannot exclude a more standard logarithmic growth ($\alpha = 1$), this behavior agrees with the RG studies of the dynamics in the spin-1/2 XXZ model with bond disorder~\cite{Altman2013}.

\begin{figure}[tb]
    \centering
    \includegraphics[width=0.99\linewidth]{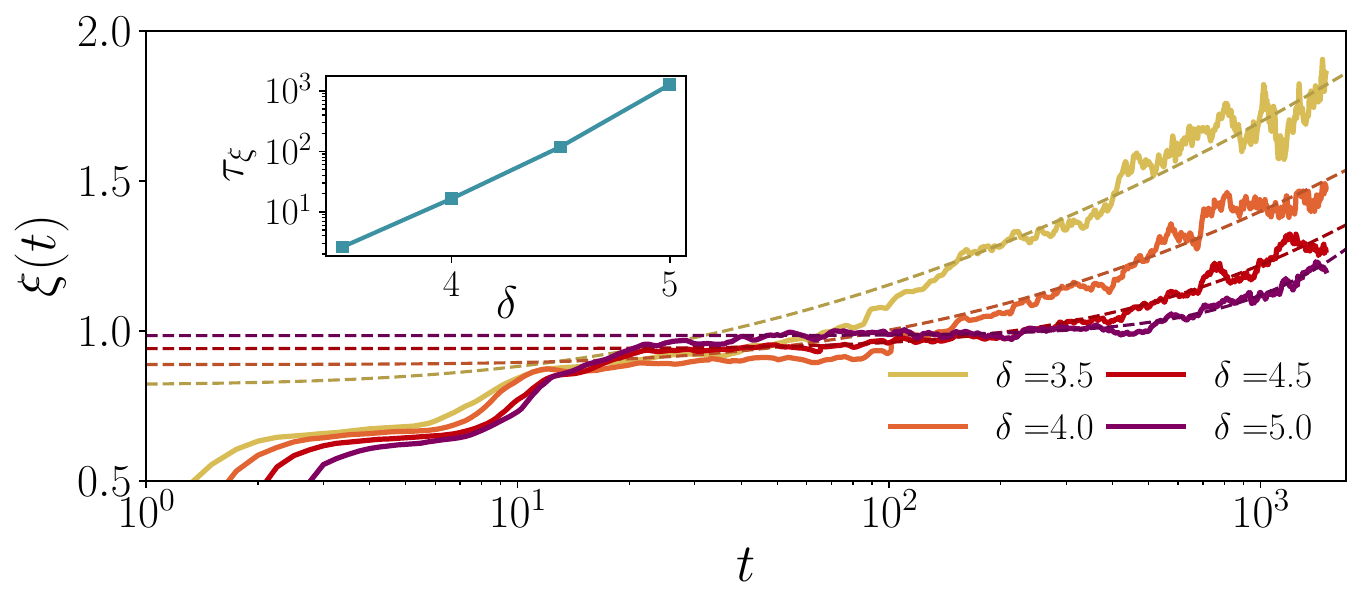}
    \caption{\label{Fig:xi}
    At short times $t<\tau_\xi$ the correlation length $\xi(t)$ reaches a plateau at a value $\xi_0\approx1$.
    On longer timescales $t>\tau_\xi$, however, its slow growth is accurately captured by a $\log^2(t)$ behavior (dashed lines), in agreement with the predictions of Ref.~\cite{Altman2013}.
    As we show in the inset, $\tau_\xi$ grows exponentially with $\delta$, thus the onset of the increase in correlation length is delayed for a time exponentially long in $\delta$.
    }
\end{figure}

\noindent\textit{Correlation length.---}
To additionally characterize the fate of the MBL spin-glass regime, we investigate the behavior pf the absolute value of the connected spin-spin correlation function averaged over disorder realizations and initial states, $\overline{|\langle\sx{i}\sx{j}\rangle|}=\overline{|\bra{\psi(t)}\sx{i}\sx{j}\ket{\psi(t)} - \bra{\psi(t)}\sx{i}\ket{\psi(t)}\bra{\psi(t)}\sx{j}\ket{\psi(t)}|}$.
We observe that the averaged correlation function decays exponentially with distance, $\overline{|\langle\sx{i}\sx{j}\rangle|}\propto \exp(-|i-j|/\xi(t))$, defining a time-dependent correlation length $\xi(t)$.
By fitting the correlation function of the central spin, $i=N/2$, to an exponential decay with distance $j$, we extract the time-dependent correlation length $\xi(t)$ and examine its evolution for different values of $\delta$ (see~\cite{Supplementary} for fit examples).

We distinguish two different behaviors of the correlation length, happening on distinct timescales.
At short times, the correlation length rapidly grows and reaches a plateau at $\xi_0\approx1$.
However, on longer timescales the correlation length resumes  its growth.
Inspired by the entanglement dynamics, we fit this growth to 
\begin{equation}
    \label{Eq:xit}
    \xi(t) = \xi_0 + a\log^2\left(\frac{t}{\tau_\xi}+1\right),
\end{equation}
corresponding to the dashed curves in Fig.~\ref{Fig:xi}, and find good agreement with the numerical data (the behavior of $\xi(t)$ could be also consistent with a power-law increase in time, see~\cite{Supplementary}).

From the fit, we extract $\tau_\xi$, and observe it grows exponentially with $\delta$, indicating an exponentially long plateau at $\xi(t)\approx\xi_0$.
It is instructive to compare this timescale to the time $\tau_S$ that governs the onset of entanglement growth beyond oscillations.
As we show in the supplementary~\cite{Supplementary}, the entanglement timescale features a parametrically slower increase with $\delta$, suggesting that at large disorders entanglement starts growing on top of the oscillations and only at a much later time the correlation length increases.

\textit{Discussion --} We studied the dynamics from random product states in the MBL-SG phase, revealing  characteristic short-time oscillations explained analytically.
Over longer timescales, TEBD simulations reveal the emergence of longer-range correlations witnessed by entanglement and spin-spin correlation functions.
Understanding the timescales of their growth, and describing them quantitatively, remains an interesting theoretical problem. 

More broadly, the slow growth of entanglement we observe is qualitatively consistent with the dynamical RG treatment constructed for a different spin-1/2 model and a specific Ne\'el initial state~\cite{Altman2013}.
Although our analytical results can serve as a starting point for extending the dynamical RG of Ref.~\cite{Altman2013} to our model, this remains an open question for future work.
Our work shows that phenomenological signatures of an infinite randomness fixed point can be detected in quench dynamics.
Extending these to the direct observation of clusters of decimated spins as inert degrees of freedom entangled among themselves but decoupled from the rest of the system, and of their hierarchical structure is a fascinating direction for further work~\cite{Pal2023}.
Finally, our results show that the MBL-SG phase can also be dynamically probed in quantum simulation platforms, since non-trivial dynamics arise at sufficiently early times.  

\begin{acknowledgments}
    \textit{Acknowledgments --}
    We thank D.~A.~Abanin for insightful discussions in the early stages of this work. 
    P.~B.~acknowledges support by the Austrian Science Fund (FWF) [Grant Agreement No.~10.55776/ESP9057324]. 
    This research was funded in whole or in part by the Austrian Science Fund (FWF) [10.55776/COE1].
    The authors acknowledge support by the European Research Council (ERC) under the European Union's Horizon 2020 research and innovation program (Grant Agreement No.~850899).
    M.~L. acknowledges support by the Deutsche Forschungsgemeinschaft (DFG, German Research Foundation) under Germany’s Excellence Strategy – EXC-2111 – 390814868. 
    The Authors acknowledge PRACE for awarding access to Joliot-Curie at GENCI@CEA, France, where the TEBD simulations were performed. 
    The TEBD simulations were performed using the ITensor library~\cite{itensor}. 
    
\end{acknowledgments}

\clearpage
\newpage
\onecolumngrid
\appendix
\begin{center}
    \textbf{Supplementary Material for ``\mytitle''}
    
    Pietro Brighi$^1$, Marko Ljubotina$^{2,3,4}$ and Maksym Serbyn$^2$
    
    \small{\textit{$\text{ }^1$Faculty of Physics, University of Vienna, Boltzmanngasse 5, 1090 Vienna, Austria}}\\
    \small{\textit{$\text{ }^2$Institute of Science and Technology Austria, am Campus 1, 3400 Klosterneuburg, Austria}}\\
	\small{\textit{$\text{ }^3$Technical University of Munich, TUM School of Natural Sciences, Physics Department, 85748 Garching, Germany}}\\
	\small{\textit{$\text{ }^4$Munich Center for Quantum Science and Technology (MCQST), Schellingstr. 4, München 80799, Germany}}
\end{center}

\renewcommand{\theequation}{S\arabic{equation}}
\setcounter{equation}{0}
\renewcommand{\thefigure}{S\arabic{figure}}
\setcounter{figure}{0}
\setcounter{page}{1}

\section{Local magnetization dynamics}

Here, we evaluate the time-evolution of local magnetization deep in the spin-glass phase $d\gg 1$.
For this purpose, we approximate the Hamiltonian neglecting the terms along the $z$ direction.
This is well motivated up to timescales $\tau\approx 1/W_h^2\gg 1$, when the effects of the terms along $z$ become relevant to the dynamics.
The evolution of the local magnetization is given by 
\begin{equation}
    \label{Eq:sz supp 1}
    \begin{split}
    \langle\sz{\ell}(t)\rangle &= \bra{\psi_0}e^{\imath\left( \sum_i J_i\sx{i}\sx{i+1} + \lambda W_J\sx{i}\sx{i+2} \right)t}\sz{\ell}e^{-\imath\left( \sum_i J_i\sx{i}\sx{i+1} + \lambda W_J\sx{i}\sx{i+2} \right)t}\ket{\psi_0} \\ 
    &= \bra{\psi_0}e^{\imath\left[J_{\ell-1}\sx{\ell-1}\sx{\ell} + J_{\ell}\sx{\ell}\sx{\ell+1} + \lambda W_J\left(\sx{\ell-2}\sx{\ell} + \sx{\ell}\sx{\ell+2} \right) \right]t}\sz{\ell}e^{-\imath\left[J_{\ell-1}\sx{\ell-1}\sx{\ell} + J_{\ell}\sx{\ell}\sx{\ell+1} + \lambda W_J\left(\sx{\ell-2}\sx{\ell} + \sx{\ell}\sx{\ell+2} \right) \right]t}\ket{\psi_0}.
    \end{split}
\end{equation}
Now, as $e^{\imath \sx{j}\sx{k} t} = \cos(t/4) + 4\imath \sx{j}\sx{k}\sin(t/4)$, one can rewrite the exponentials in the equation above as a sum of sines and cosines
\begin{equation}
    \label{Eq:sz supp 2}
    \begin{split}
    \langle&\sz{\ell}(t)\rangle = \bra{\psi_0}\left(\cos\left(\frac{J_{\ell-1}}{4}t\right) + 4\imath \sx{\ell-1}\sx{\ell}\sin\left(\frac{J_{\ell-1}}{4}t\right) \right) \left( \cos\left(\frac{J_{\ell}}{4}t\right) + 4\imath \sx{\ell}\sx{\ell+1}\sin\left(\frac{J_{\ell}}{4}t\right) \right) \\
    &\left( \cos\left(\frac{\lambda W_J}{4} t\right) + 4\imath \sx{\ell-2}\sx{\ell}\sin\left(\frac{\lambda W_J}{4} t\right)\right) \left(\cos\left(\frac{\lambda W_J}{4} t\right) + 4\imath \sx{\ell}\sx{\ell+2}\sin\left(\frac{\lambda W_J}{4}t\right) \right) \sz{\ell} \\ 
    &\left(\cos\left(\frac{J_{\ell-1}}{4}t\right) - 4\imath \sx{\ell-1}\sx{\ell}\sin\left(\frac{J_{\ell-1}}{4}t\right) \right) \left( \cos\left(\frac{J_{\ell}}{4}t\right) - 4\imath \sx{\ell}\sx{\ell+1}\sin\left(\frac{J_{\ell}}{4}t\right) \right) \\
    &\left( \cos\left(\frac{\lambda W_J}{4} t\right) - 4\imath \sx{\ell-2}\sx{\ell}\sin\left(\frac{\lambda W_J}{4} t\right)\right) \left(\cos\left(\frac{\lambda W_J}{4} t\right) - 4\imath \sx{\ell}\sx{\ell+2}\sin\left(\frac{\lambda W_J}{4}t\right) \right)\ket{\psi_0}.
    \end{split}
\end{equation}
Since $\ket{\psi_0}$ is a product state in the $\sz{}$ basis, all terms involving Pauli matrices acting on different sites on the left and on the right of $\sz{\ell}$ identically vanish.
Therefore, one remains with
\begin{equation}
    \begin{split}
        \label{Eq:sz supp 3}
        \langle\sz{\ell}(t)\rangle &= \langle\sz{\ell}\rangle_0 \Biggr\{ \cos^4\left(\frac{\lambda W_J}{4} t\right)\left[ \cos^2\left(\frac{J_{\ell}}{4}t\right) - \sin^2\left(\frac{J_{\ell}}{4}t\right)\right]\left[ \cos^2\left(\frac{J_{\ell-1}}{4}t\right) - \sin^2\left(\frac{J_{\ell-1}}{4}t\right) \right] \\
        & - 2\cos^2\left(\frac{\lambda W_J}{4}t\right)\sin^2\left(\frac{\lambda W_J}{4}t\right)\left[ \cos^2\left(\frac{J_{\ell}}{4}t\right) - \sin^2\left(\frac{J_{\ell}}{4}t\right)\right]\left[ \cos^2\left(\frac{J_{\ell-1}}{4}t\right) - \sin^2\left(\frac{J_{\ell-1}}{4}t\right) \right] \\
        & + \sin^4\left( \frac{\lambda W_J}{4}t \right)\left[ \cos^2\left(\frac{J_{\ell}}{4}t\right) - \sin^2\left(\frac{J_{\ell}}{4}t\right)\right]\left[ \cos^2\left(\frac{J_{\ell-1}}{4}t\right) - \sin^2\left(\frac{J_{\ell-1}}{4}t\right) \right] \Biggr\} \\
        &= \langle \sz{\ell}\rangle_0 \left[ \cos^2\left(\frac{J_{\ell}}{4}t\right) - \sin^2\left(\frac{J_{\ell}}{4}t\right)\right]\left[ \cos^2\left(\frac{J_{\ell-1}}{4}t\right) - \sin^2\left(\frac{J_{\ell-1}}{4}t\right) \right]\left[ \cos^2\left(\frac{\lambda W_J}{4}t\right) - \sin^2\left(\frac{\lambda W_J}{4}t\right) \right]^2 \\
        & =  \langle \sz{\ell}\rangle_0 \cos\left( \frac{J_\ell}{2}t \right)\cos\left( \frac{J_{\ell-1}}{2}t \right)\cos^2\left( \frac{\lambda W_J}{2}t \right),
    \end{split}
\end{equation}
which is Eq.~(\ref{Eq:szt}) in the main text.

The analytical expression above further allows for exact disorder averaging of the magnetization dynamics.
As each disorder realization $(n)$ introduces an independent set of frequencies, determined by the random couplings $J^{(n)}_\ell$, the exact disorder average can be obtained by integrating Eq.~(\ref{Eq:sz supp 3}) over the distribution function generating the random variables.
In the case of a box distribution with unit weight studied here, that corresponds to the characteristic function of the interval $[0,W_J]$, with $W_J=1$ in the spin-glass phase.
The average magnetization can be then obtained as
\begin{equation}
    \overline{\langle  \sz{\ell}(t)\rangle} = \langle\sz{\ell}\rangle_0\cos^2\left( \frac{\lambda W_J}{2}t \right)\left(\int_0^{W_J} dJ \cos\left(\frac{J}{2}t\right) \right)^2 = 4\langle\sz{\ell}\rangle_0\frac{\cos^2\left( \frac{\lambda W_J}{2}t \right)\sin^2\left(\frac{W_J}{2} t\right)}{t^2},
\end{equation}
corresponding to Eq.~(\ref{Eq:sz average}) in the main text.

\section{MPO expression for the time propagator}

\begin{figure}[t]
    \centering\includegraphics[width=0.75\linewidth]{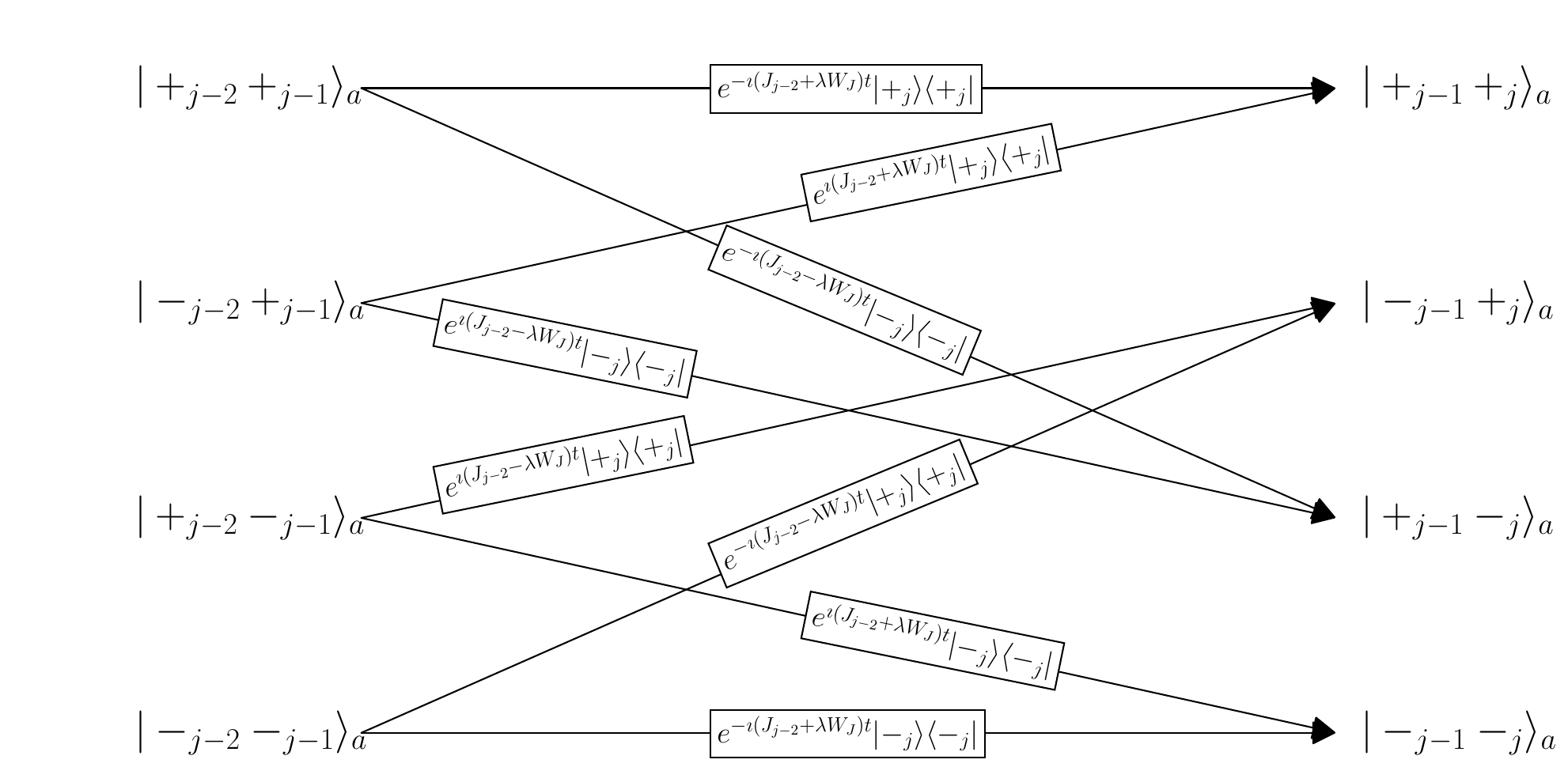}
    \caption{\label{Fig:MPO scheme}
    Automaton representation of the action of the MPO.
    The auxiliary space encodes information about the state of the two sites to the left of a given spin.
    This allows to give the correct signs to the terms in the exponent.
    }
\end{figure}

Deep in the spin-glass phase $\delta\gg 1$, the time propagator has a particularly simple structure, which allows us to write it as a low bond dimension matrix product operator (MPO), as mentioned in the main text.
Here, we show how the MPO is obtained.

The Hamiltonian Eq.~(\ref{Eq:H}) can be written separating terms diagonal in the $z$ and $x$ bases
\begin{equation}
    \label{Eq:H suppl}
    H = H_{W_J} + H_{W_h} = \underbrace{\sum_i\left( J_i\sx{i}\sx{i+1} +  \lambda W_J\sx{i}\sx{i+2}\right)}_{H_{W_J}}+ \underbrace{\sum_i \left(h_i\sz{i}  +\lambda W_h\sz{i}\sz{i+1}\right)}_{H_{W_h}}.
\end{equation}
In the spin-glass phase $W_h\ll1$, and we can approximate $H\approx H_{W_J}$.
Since all terms in the $H_{W_J}$ Hamiltonian commute with one another we can write the time propagator as a product of local operators acting on three sites: $U_t = e^{-\imath t \sum_j h_{j-2,j-1,j}} = \prod_j e^{-\imath t  h_{j-2,j-1,j}}$, with $h_{j-2,j-1,j} = J_{j-2}\sx{j-2}\sx{j-1} + \lambda W_J\sx{j-2}\sx{j}$.
Each local term in the product is diagonal in the $x$ basis, with its coefficients determined by the state of the two spins to its left.

One can conveniently write the time propagator $U_t$ as a bond dimension $\chi=4$ MPO $U_t = \prod_j A^{(j)}_t$ with
\begin{equation}
    \label{Eq:Ut MPO}
    A^{(j)}_t = \begin{pmatrix}
        & e^{-\imath(J_{j-2} + \lambda W_J)t}\ket{+_j}\bra{+_j} & 0 & e^{-\imath(J_{j-2} - \lambda W_J)t}\ket{-_j}\bra{-_j} & 0 \\
        & e^{\imath(J_{j-2} + \lambda W_J)t}\ket{+_j}\bra{+_j} & 0 & e^{\imath(J_{j-2} - \lambda W_J)t}\ket{-_j}\bra{-_j} & 0 \\
        & 0 & e^{\imath(J_{j-2} - \lambda W_J)t}\ket{+_j}\bra{+_j} & 0 & e^{\imath(J_{j-2}+ \lambda W_J)t}\ket{-_j}\bra{-_j} \\
        & 0 & e^{-\imath(J_{j-2} - \lambda W_J)t}\ket{+_j}\bra{+_j} & 0 & e^{-\imath(J_{j-2} + \lambda W_J)t}\ket{-_j}\bra{-_j} \\
    \end{pmatrix}.
\end{equation}
As mentioned above, at each site $j$ the sign of the coefficients in the exponential is determined by the state of the two spins to the left of $j$.
Therefore, the auxiliary dimension of the MPO has to encode the information on the two sites to the left, thus leading to $\chi=4$ corresponding of the possible states of $j-1$ and $j-2$ in the $x$ basis, in which the operators are diagonal. 
A schematic automaton picture~\cite{Bacon2008} of the MPO is represented in Figure~\ref{Fig:MPO scheme}.

The explicit construction of the MPO defining the time evolution is useful in obtaining relevant features of the entanglement entropy.
First, as $U_t$ has a finite bond dimension $\chi=4$ the entanglement entropy cannot exceed $S = \log(\chi) = 2\log(2)$, as shown in the main text.
Furthermore, from the analysis of the exponent we notice that there are two terms contributing to the dynamics: a random frequency $J_{j-2}$ which is different on each site and in all disorder realizations, and a fixed frequency $\lambda W_J$ independent of site, disorder realization and disorder strength since $W_J = 1$ as $\delta>0$.
In the average entanglement dynamics, then, the random contributions will typically average out, leaving a single dominant frequency for the oscillations at $\omega_0 = \lambda W_J$.
Both these predictions are confirmed by comparison with numerical simulations shown in the main text.

\section{Examples of correlation function exponential fit and power law fit for $\xi(t)$}

In the main text we  defined the corelation length as the length scale for the exponential decay of the disorder averaged absolute value of the correlation function $\overline{|\langle \sx{i}\sx{j }\rangle|}\propto e^{-|i-j|/\xi}$.
We notice that for our particular choice of initial state $\bra{\psi(t)}\sx{j}\ket{\psi(t)} = 0,\;\forall j,t$ and therefore the connected part of the correlation function coincides with $\langle \sx{i}\sx{j }\rangle$.
In an infinite system, the disorder averaged correlation length is expected to be the same on all sites, as each local disorder is independently drawn.
However, as we deal with finite systems, we chose to study the decay of the correlation function from the central spin $i=N/2$ to reduce the finite size effects.
This defines the correlation length $\xi_{N/2}$ shown in the main text, which upon sufficient disorder averaging coincides with the generic correlation length $\xi$.

\begin{figure}[h]
    \centering
    \includegraphics[width=0.99\linewidth]{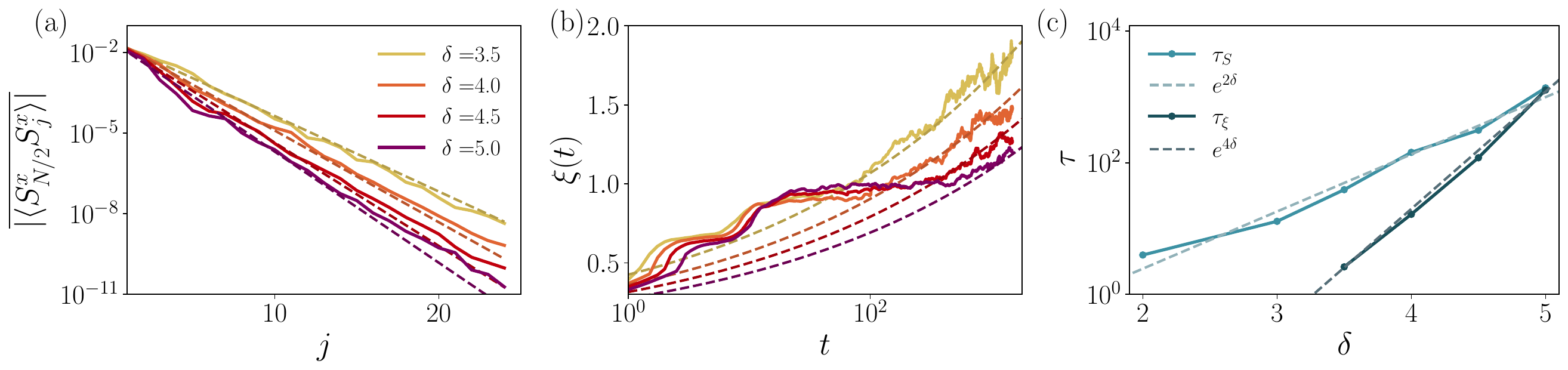}
    \caption{\label{Fig:Exp fit and alternative PL fit}
    (a): The correlation functions at time $t=100$ for all the values of $\delta$ within the spin glass phase show a clear exponential decay. 
    The dashed lines represent the fit with the corresponding $\xi_{N/2}(t)$.
    (b): Alternative fit of the correlation length to a power law $\propto t^{0.2}$ (dashed lines).
    The power law fit is in good agreement with the numerical data, but it is not supported by any theoretical prediction.
    (c): Comparison of the two timescales $\tau_\xi$ and $\tau_S$ shows a parametrically slower increase of $\tau_S$ with $\delta$.
    While in this parameter regime $\tau_\xi\leq\tau_S$, asymptotically entanglement entropy would start growing much earlier than the correlation length.
    }
\end{figure}

In this Section, we show numerical data confirming the exponential decay of the disorder averaged correlation function, as shown in Figure~\ref{Fig:Exp fit and alternative PL fit}~(a).
Our numerical simulations show a clear exponential behavior (modulo small oscillations due to finite disorder average) in good agreement with the fit (dashed lines).
The data are shown for a single arbitrary time $t=100$, but do not change qualitatively for different choices of $t$.

The correlation length $\xi(t)$ was fitted in the main part to the square logarithmic behavior predicted for the entanglement entropy by renormalization group analysis~\cite{Altman2013}.
However, the range of growth of the correlation is too limited to uniquely determine its functional behavior.
Here, we report an alternative power law fit, whose precision is comparable to the one used in the main text (although it is not supported by any theoretical argument).
In Figure~\ref{Fig:Exp fit and alternative PL fit}~(b) we show the correlation length dynamics for all the values of $\delta$ within the MBL-SG phase together with a power law fit (dashed lines).
The algebraic growth $\propto t^{0.2}$ is similar for all values of $\delta$, with a coefficient $\mathcal{A}(\delta)$ decaying with $\delta$.
Both the power law and the logarithmic behavior shown in the main text define a timescale for the growth of the correlation length beyond $\xi_0$ growing exponentially with $\delta$.
In Figure~\ref{Fig:Exp fit and alternative PL fit}~(c) we compare this timescale $\tau_\xi$ as obtained from Eq.~(\ref{Eq:xit}) with the analogue obtained for entanglement growth.
Remarkably, the two have different scaling with $\delta$, with $\tau_S$ increasing more slowly than $\tau_\xi$.
While asymptotically then entanglement would start growing earlier than correlations, within this range of parameters $\tau_\xi\leq\tau_S$, thus implying a faster growth of the correlation length as compared with entanglement entropy.

\section{Bond dimension analysis}

In this Section we present the bond dimension analysis of our tensor network simulations.
In particular, we focus on the spin-glass region of the parameter space, $\delta\geq3.5$, which is the central topic of the main part of the text.

In matrix product states (MPS), as the wavefunction gets decomposed in $N$ order-$3$ tensors, one defines a truncation value $\varepsilon$ such that all singular values $s_n<\varepsilon$ are neglected, and a maximum bond dimension $\chi$ corresponding to the maximum number of singular values that can be stored per bond.
Therefore, to ensure the accuracy of numerical results, one needs to check the convergence in either of these parameters.

\begin{figure}
    \centering
    \includegraphics[width=0.99\linewidth]{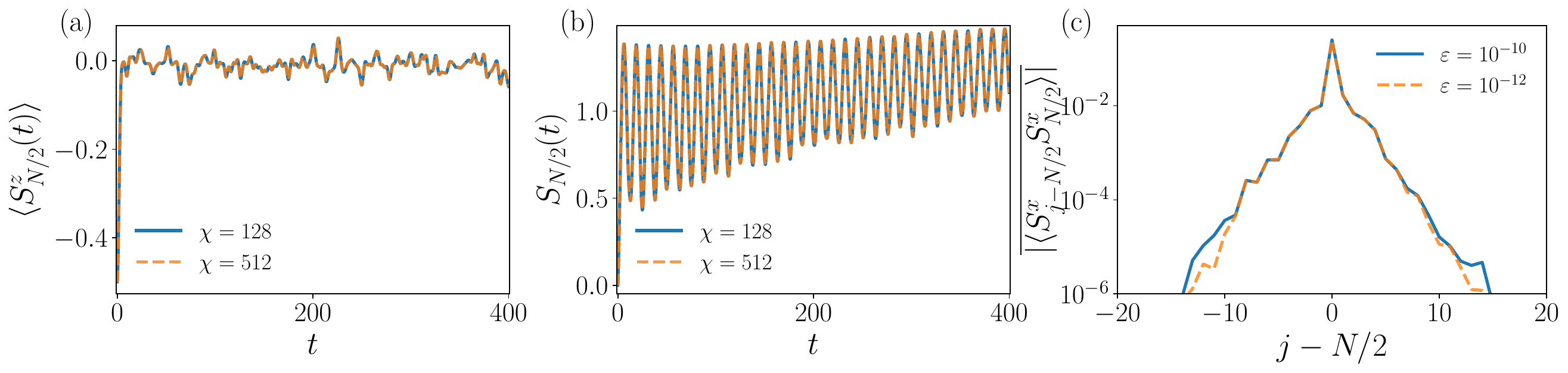}
    \caption{\label{Fig:chi comparison}
    Bond dimension comparison for local magnetization~(a) and entanglement entropy~(b) at the central site $\ell = N/2$ and for $\delta=3.5$ (worst case for $\chi$ convergence as it is the closest to the ergodic region).
    The curves for the two bond dimensions nicely overlap, confirming the convergence of the numerical simulations.
    For the correlation functions~(c) at $\delta=3.5$ we compare the exponential decay for two different truncation thresholds at $\chi=512$, where the MPS does not saturate the bond dimension.
    At late times $t=400$ the two curves overlap up to $\mathcal{O}(10^{-4}$).
    }
\end{figure}

In Figure~\ref{Fig:chi comparison}, we compare the observables studied in the main text.
In panels~(a) and~(b), we show local magnetization and entanglement entropy in the center of the chain at $\delta=3.5$ and for two values of bond dimension $\chi=128,512$.
The curves corresponding to the two different bond dimensions perfectly overlap, thus ensuring convergence of our numerical simulations.
In panel~(c), instead, we show the behavior of the correlation function for two different values of the truncation threshold at fixed $\chi=512$ and at time $t=400$.
As this bond dimension never gets saturated at these values of $\varepsilon$ (i.e.~the number of singular values is smaller than $\chi$), it is worth comparing two different truncation thresholds, especially for a quantity like the correlation function, where we care about its exponential decay.
As the figure shows, the curves overlap perfectly up to values $\mathcal{O}(10^{-4})$, thus again ensuring the reliability of our data.

\end{document}